\documentclass[aps,prb,twocolumn,groupedaddress,showkeys,showpacs]{revtex4-1}

\usepackage{graphicx}
\usepackage{dcolumn}
\usepackage{bm}
\usepackage{amsmath}
\usepackage{amssymb}
\usepackage{latexsym}
\usepackage{epsfig}
\usepackage{amsbsy}
\usepackage{array}
\usepackage{amssymb}
\usepackage{setspace}
\usepackage{bm}
\usepackage{color}

\begin{document}

\title{Enhanced superconductivity in hole-doped Nb$_{2}$PdS$_{5}$}

\author{Qian Chen$^{1,3}$}
\author{Xiaohui Yang$^{1,3}$}
\author{Xiaojun Yang$^{1,3}$}
\author{Jian Chen$^{1,3}$}
\author{Chenyi Shen$^{1,3}$}
\author{Pan Zhang$^{1,3}$}
\author{Yupeng Li$^{1,3}$}
\author{Qian Tao$^{1,2}$}

\author{Zhu-An Xu$^{1,2,3}$}
\email{zhuan@zju.edu.cn}

\affiliation{$^1$Department of Physics and State Key Laboratory of Silicon Materials, Zhejiang University, Hangzhou 310027, China}
\affiliation{$^2$Zhejiang California International NanoSystems Institute, Zhejiang University, Hangzhou 310058, China}
\affiliation{$^3$Collaborative Innovation Centre of Advanced Microstructures, Nanjing 210093, China}

\date{\today}

\begin{abstract}
We synthesized a series of Nb$_{2}$Pd$_{1-x}$Ru$_{x}$S$_{5}$
polycrystalline samples by a solid-state reaction method and
systematically investigated the Ru-doping effect on
superconductivity by transport and magnetic measurements. It is
found that superconductivity is enhanced with Ru doping and is quite
robust upon disorder. Hall coefficient measurements indicate that
the charge transport is dominated by hole-type charge carriers
similar to the case of Ir doping, suggesting multi-band
superconductivity. Upon Ru or Ir doping, \emph H$_{c2}$/$\emph
T_c$ exhibits a significant enhancement, exceeding the Pauli
paramagnetic limit value by a factor of approximately 4. A comparison
of $T_c$ and the upper critical field ($H_{c2}$) amongst the different
doping elements on Pd site, reveals a significant role of spin--orbit
coupling.
\end{abstract}


\maketitle

\section{Introduction}

The transition metal-chalcogenide compounds T$_{2}$PdCh$_{5}$, where
\emph T = Nb or Ta and \emph {Ch} = S or Se, are
quasi-one-dimensional (Q1D) superconductors with a remarkably high
upper critical field\cite{Nb2Pd0. 81S5,band
structure,Ta2PdSe5,Ta2PdS5}, which surpasses by far the expected
Pauli limiting field (\emph H$_{c2}^{Pauli}$ = 1.84$\emph
T_c$)\cite{pauli}. These compounds achieving superconducting
temperatures up to 6 K are proposed to be multi-band
superconductors\cite{Nb2Pd0. 81S5,Ta2PdSe5,multi,calculation}. Band structure calculations\cite{Nb2Pd0. 81S5,band
structure, calculation} show that the Fermi surface of
Nb$_{2}$PdS$_{5}$ is composed of multiple sheets, i.e.,
two-dimensional sheets with hole character and Q1D sheets with both
electron and hole character, and this system may be in proximity
to a magnetically- or charge density wave- (CDW-) ordered state, owing to
the nesting properties of those Q1D Fermi surface sheets. The
increase in the $d$ electron population on the Pd site is expected to
flatten the Q1D Fermi surface sheets; thus, it may enhance the
nesting properties and an unconventional
superconducting pairing scenario is suggested\cite{Nb2Pd0.
81S5,calculation}. It is shown that partial substitution of Pd by
Ni\cite{Ni doping} or Ir\cite{Ir doping} leads to slightly
enhanced \emph T$_c$, but superconductivity is suppressed in the
cases of Pt-for-Pd doping \cite{Ni doping} or Ag-for-Pd
doping\cite{Ir doping}, and Se-for-S doping \cite{Se doping}.
However, the feature of large \emph H$_{c2}$ relative to $\emph
T_c$ is found to be robust against these substitutions, providing strong experimental evidence of unconventional superconductivity
in this system.
\begin{figure*}
\includegraphics[width=5.5in]{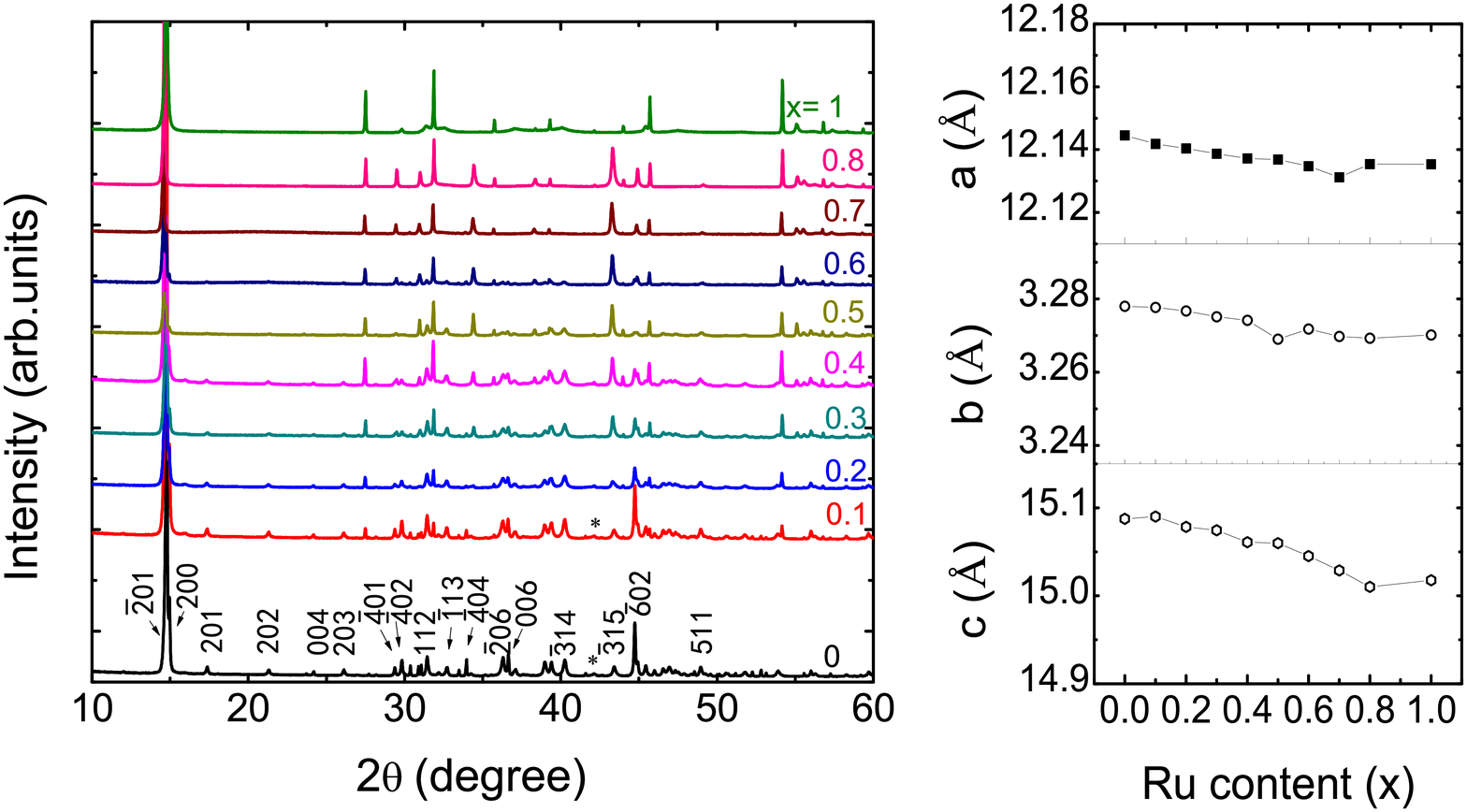}\hspace{5pc}
\begin{center}
\caption {\label{fig1}Left panel: room temperature powder X-ray
diffraction patterns for the Nb$_{2}$Pd$_{1-x}$Ru$_{x}$S$_{5}$
samples with \emph x= 0$\sim$1. Right panel: variations of lattice
constants as a function of nominal doping content $x$. }
\end{center}
\end{figure*}

As is often the case, high superconducting critical field can be
attributed to the multi-band effect\cite{multi-band},
strong-coupling\cite{strong-coupling}, spin--triplet
pairing\cite{triplet}, or strong spin--orbit coupling
(SOC)\cite{spin-orbit}. A study on the effect of selenium doping
on Nb$_{2}$PdS$_{5}$ rules out the strong-coupling as well as
spin--triplet pairing as the origin of the exotic
superconductivity\cite{Se doping}. In contrast, it is suggested
that the large \emph H$_{c2}$ can be attributed to the strong SOC
associated with the heavy Pd element\cite{Ta2PdS5,Ni
doping, strong-coupling, spin-orbit}. Moreover, electronic structure
calculations for Ta$_2$PdS$_5$ show that the large $H_{c2}$ is a
result of a combination of strong coupling and multi-band effects
in the extreme dirty limit\cite{calculation}. These experimental
and theoretical studies have verified the importance of the
presence of Pd irons with a large Z number to the unconventional
properties, where Z is the atomic mass number. Recently, it has been reported that the charge carrier
density (or band filling), which can be modulated by
hole(electron)-type doping, could be a crucial factor for tuning
superconductivity in this system\cite{Ir doping}. So far, the
origin of the large \emph H$_{c2}$ and exotic superconductivity in
T$_{2}$PdCh$_{5}$ has not been understood.

In this paper, we report on a detailed investigation of the hole-type
Ru doping effect on superconductivity and compare with the case of
similar Ir doping. Both Ru and Ir doping on Pd site can be
considered as a hole-type doping, but less heavy Ru should have
a weaker SOC compared with Ir or Pt doping; thus, it may help with
distinguishing the different effects of charge carrier density and
SOC upon superconductivity. We make a comparison of the cases
of Ir, Ag, Ni, and Pt doping and it turns out that $T_c$ and \emph
H$_{c2}$ can be slightly enhanced by hole-type doping such as Ru
or Ir doping, but that Ir doping has a more significant effect owing to its
stronger SOC. Our work implies that there could be a close
relationship between the extremely large \emph H$_{c2}$ and SOC in
this Q1D superconducting system.

\section{Experimental details}
We synthesized a series of Nb$_{2}$Pd$_{1-x}$Ru$_{x}$S$_{5}$
(\emph x = 0--1 ) and Nb$_{2}$Pd$_{1-x}$Ir$_{x}$S$_{5}$ (\emph
x = 0.1, 0.2, 0.4) polycrystalline samples by the usual solid-state
reaction method. More details can be found in our previous
report\cite{Ir doping}. Nb, Pd, Ir or Ru, and S of high purity
(99.9$\%$) were used as the starting materials. Energy-dispersive
X-ray(EDX) spectroscopy analysis shows the presence of Pd-site
deficiencies in this system, which are found to be about 0.25,
similar to our previous report. The actual Ru composition of the
obtained samples is in proportion to the nominal stoichiometric
composition, and 70--80\% of the corresponding nominal Ru
element is doped into the grains, as determined by the EDX
analysis. Powder X-ray diffraction (XRD) measurements were
performed at room temperature on a PANalytical X-ray
diffractometer with Cu K$\alpha$ radiation and the lattice constants
were determined using the program X'Pert HighScore. The temperature
dependence of magnetization was measured on a Quantum Design
magnetic property measurement system (MPMS-5). The electrical
resistivity and Hall coefficient (\emph R$_H$) measurements were
carried out in a physical property measurement system (PPMS-9) by
a standard six-terminal method.

\section{Results and discussion}

\begin{figure}
\includegraphics[width=20pc]{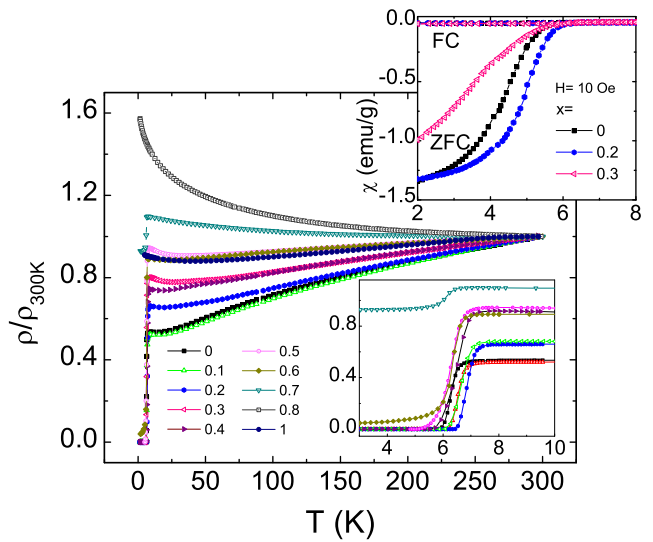}
\caption{\label{fig2}Temperature dependence of the electrical
resistivity  of Nb$_{2}$Pd$_{1-x}$Ru$_{x}$S$_{5}$ normalized to
$\rho_{300 K}$. The lower inset shows the expanded view of
resistivity versus \emph T around the superconducting transition
temperatures. The upper inset depicts temperature dependence of magnetic susceptibility under zero-field cooling (ZFC) and filed cooling (FC) for Nb$_{2}$Pd$_{1-x}$Ru$_{x}$S$_{5}$ (\emph x= 0, 0.2 and 0.3), respectively.}
\end{figure}

Fig. \ref{fig1} shows XRD patterns of
Nb$_{2}$Pd$_{1-x}$Ru$_{x}$S$_{5}$ (\emph x = 0--1). The main
peaks are well indexed based on a monoclinic cell with the space
group C2/m, indicating the samples are almost single phase. Only a
few minor impurity peaks marked by the asterisks are observed,
which are unknown yet. The lattice parameters as a function of
nominal \emph x are presented in the right panel. Both \emph a and
\emph b remain almost constant, while the \emph c-axis shrinks
significantly with increasing Ru doping. This result is consistent
with the smaller ion radius of Ru compared with that of Pd. The structural
characterization of Ir-doped Nb$_{2}$Pd$_{1-x}$Ir$_{x}$S$_{5}$
samples can be found in the previous report\cite{Ir doping}.

The temperature-dependent resistivity shown in Fig. \ref{fig2} for
all studied Ru-doped samples is normalized, whereby each value is
divided by its individual absolute value at room temperature
($\rho_{300 K}$). The resistivity at \emph T = 300 K ranges from
1.55 m$\Omega$$\cdot$cm to 4.6 m$\Omega$$\cdot$cm. For the parent
compound Nb$_{2}$PdS$_{5}$, the resistivity shows a  metallic
behavior upon decreasing temperature and then becomes
superconducting below $\emph T_c$ $\sim$ 6 K, consistent with
previous reports\cite{Nb2Pd0. 81S5, Ir doping}. Upon doping with
Ru, a small upturn shows up at a low temperature, which can be
interpreted as the result of Anderson localization\cite{Ta2PdS5}
or the grain boundary effect\cite{grainboundary}. In the scenario of
disorder-induced localization, a more profound resistivity upturn
would be observed with the increase of doping. An expanded plot of
the low-temperature regime below 10 K is presented in the lower inset
of Fig. \ref{fig2}. Remarkably, the increase in Ru doping enhances
superconductivity and $\emph T_c^{mid}$ reaches 6.9 K at \emph x =
0.2, and then superconductivity is suppressed with further doping.
In addition, we confirmed the bulk nature of superconductivity by
the magnetic susceptibility measurements for all samples. The data
for three samples (\emph x = 0, 0.2,and 0.3) are presented in the
upper right inset of Fig.2. Strong diamagnetic signals were
observed, and meanwhile the onset temperature of magnetic
transitions obviously varies with the Ru content, which is consistent with
the resistivity measurements.

\begin{figure}[h]
\includegraphics[width=20.4pc]{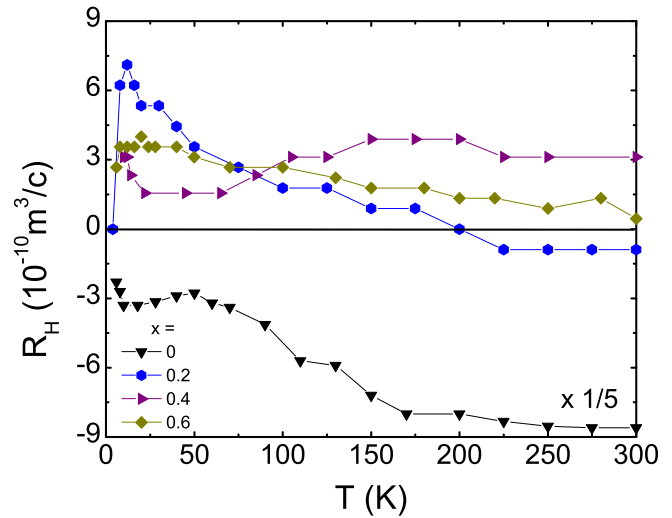}
\caption{\label{fig3} Temperature dependence of the Hall
coefficient for the four superconducting samples (\emph x= 0, 0.2,
0.4 and 0.6). The data for undoped sample have been divided by 5
for clarity.}
\end{figure}

In order to get an insight into the evolution of transport
properties, the temperature dependence of the Hall coefficient at
various doping levels was measured. As shown in Fig. \ref{fig3},
\emph R$_H$ is positive for the doped samples, indicating
hole-type dominant charge carriers, as opposed to the $n$-type charge
carriers in the parent compound. This confirms that Ru doping
induces the holes into the system. It is noted that for the
\emph x = 0.2 sample, while \emph R$_H$ is negative at
room temperature, it becomes positive and strongly \emph T
dependent below 200 K. In a scenario in which the Fermi surface
contains both electron and hole pockets, the sign change of \emph
R$_H$ as a function of temperature allows us to suggest that hole
carriers dominate the transport properties at low temperature,
owing to the different temperature dependence of mobility for
different types of charge carriers, or possible reconstruction of
the Fermi surface as reported in the
cuprates\cite{reconstruction}. As for the undoped sample, whereas
above 200 K, \emph R$_H$ is relatively insensitive to temperature,
it monotonically increases with decreasing \emph T. These data may
provide evidence for the possible multi-band superconductivity.
For conventional one-band metals, in contrast, weak temperature
dependence of normal state \emph R$_H$ is usually observed. Beyond
these considerations, it is known that in the transition metal
dichalcogenides, the change in electronic structure due to
magnetic ordering or CDW transition could lead as well to a sign
change of the Hall coefficient\cite{ sign changing}. Though the
Nb$_{2}$PdS$_{5}$ system is predicted to be in proximity to a
magnetic order or CDW state, there is no experimental evidence of this to date.

According to our previous work\cite{Ir doping}, partial
substitution of Pd by Ir, which is also considered as hole-type
doping, could obviously increase \emph T$_c$ for low Ir content,
while Ag doping, which is regarded as electron-type doping, destroys
superconductivity quickly. It is interesting to compare the effects
of Ru doping with Ir doping, as both are regarded as
hole-type dopants. Since the strength of SOC is proportional to
$Z^4$, Ru doping should hardly change the SOC associated with the Pd site,
while Ir or Pt (Ni) doping could increase (decrease) the SOC in the
system \cite{Ni doping, spin-orbit}. Therefore, we may distinguish
the effect of SOC from charge carrier density through comparing
the effect of Ru doping with that of Ir doping.

\begin{figure}[h]
\includegraphics[width=20pc]{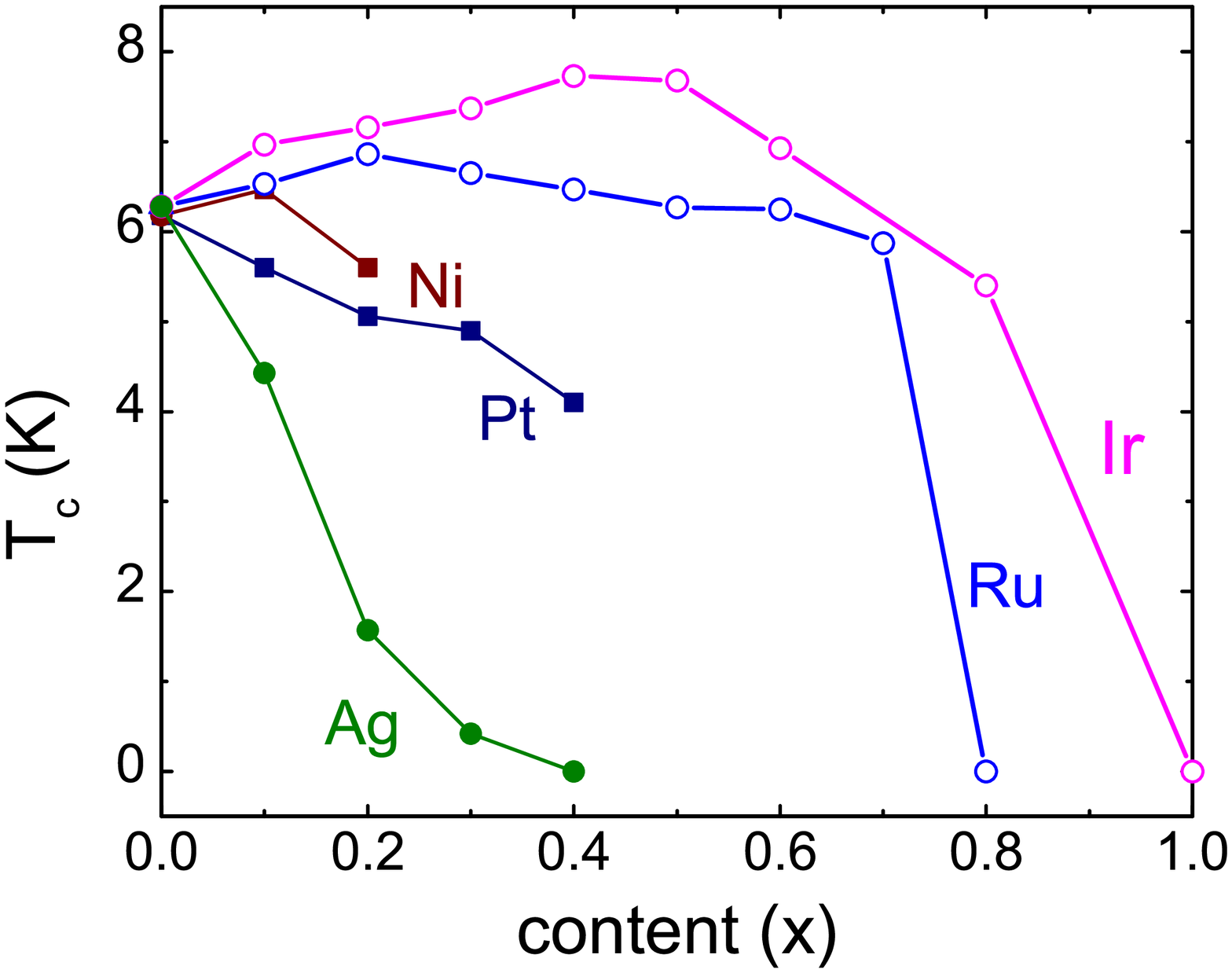}
\caption{\label{fig4}\emph T$_c$ as a function of doping level
\emph x for Nb$_{2}$Pd$_{1-x}$Ru$_{x}$S$_{5}$, extracted from the
resistivity where $\rho$(\emph T) drops to 50$\%$ of the normal
state value. The variations of $T_c$ of
Nb$_{2}$Pd$_{1-x}$R$_{x}$S$_{5}$, $R$ = Ir, Ag, Ni or Pt, is also
plotted for comparison. The data for Ni and Pt doping are taken
from Ref.\cite{Ni doping}. The data for Ir and Ag doping are from
our previous work\cite{Ir doping}. The hollow symbols stand for
hole doping, Ru and Ir, distinct from other dopings which are in
solid symbols.}
\end{figure}

\begin{figure*}
\begin{center}
\includegraphics[width=5.6in]{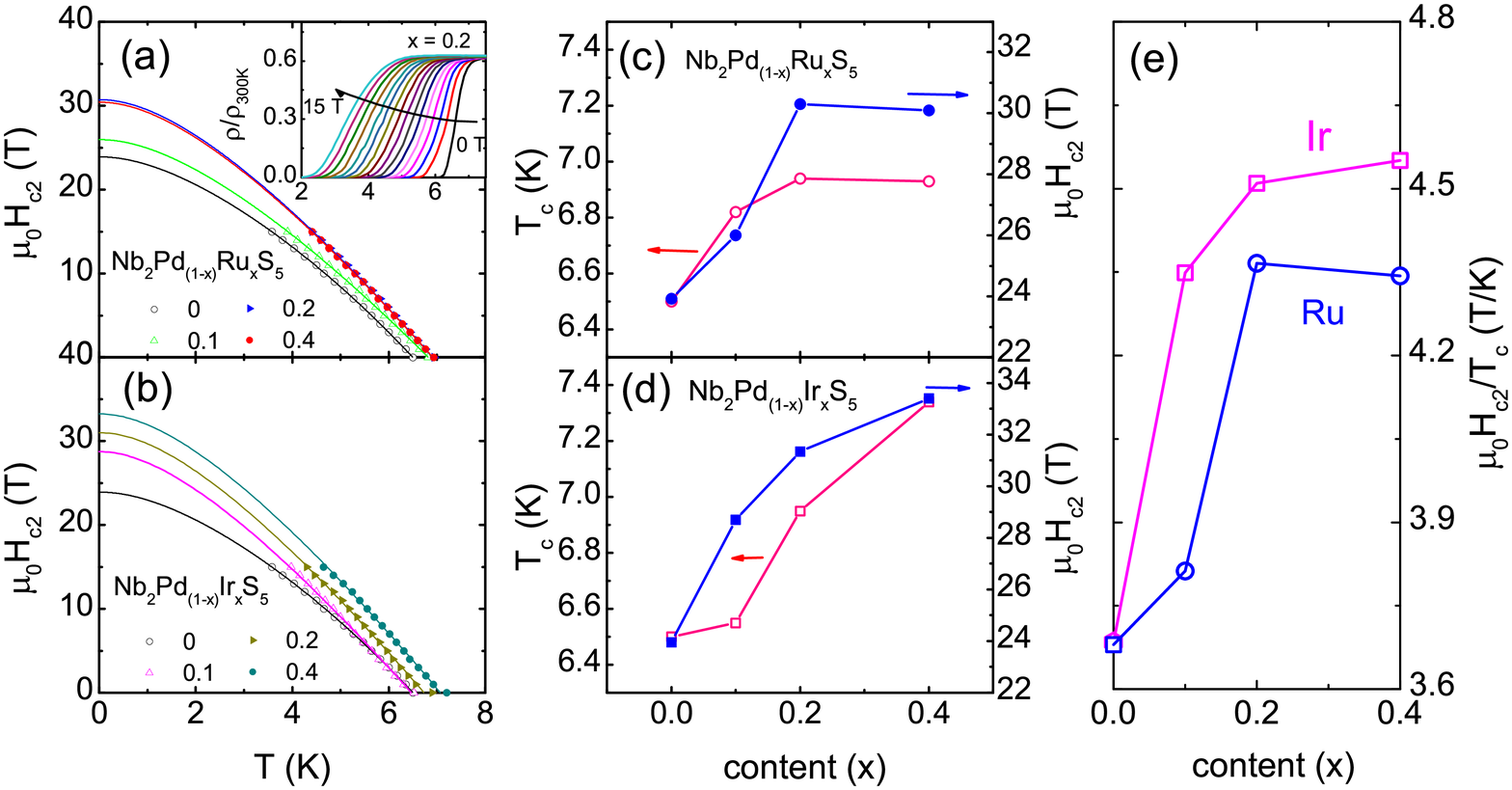}

\caption {\label{fig5}(a) and (b) Comparison of the upper critical
field data \emph H$_{c2}$ for Nb$_{2}$Pd$_{1-x}$Ru$_{x}$S$_{5}$
and Nb$_{2}$Pd$_{1-x}$Ir$_{x}$S$_{5}$. Inset: Resistivity as a
function of  temperature under several magnetic field (0$\sim$15
T). (c) and (d) The doping dependence of extrapolated \emph
H$_{c2}$(0) and \emph T$_c$ values. (e) Variations in \emph
H$_{c2}$/\emph T$_c$ values with the doping content. }
\end{center}
\end{figure*}

Based on the resistivity measurements, a phase
diagram of $T_c$ vs. doping level for the Nb$_{2}$Pd$_{1-x}$Ru$_{x}$S$_{5}$ series is constructed as shown in Fig. \ref{fig4}. The data of Ir, Ag, Pt, and Ni doping (Ref. 8 and 9) are also plotted for comparison. Ru and
Ir doping are indicated by hollow symbols standing for hole doping; Ag doping is indicated by solid
characters. It can be found
that superconductivity is enhanced with a maximum \emph
T$_c$$^{mid}$ of 6.86 K by Ru (\emph x = 0.2) doping, and 7.73 K
by Ir (\emph x = 0.4) doping. Both Ru and Ir doping
are expected to increase the hole-type carrier density, which may
drive the system far away from the magnetic order or a possible
CDW order and thus favor superconductivity, while the
electron-type doping (i.e., Ag doping) has an opposite effect. In
this scenario, we could understand why \emph T$_c$ initially
increases for both Ru doping and Ir doping.
However, the maximum
$T_c$ is even higher and the superconducting range (up to $x$
of 0.8) is wider for the Ir doping case. We propose that the
enhanced SOC strength owing to heavier Ir may account for this
difference. This result implies that not only the charge carrier
density but also SOC could play an important role in
controlling superconductivity of Nb$_{2}$PdS$_{5}$. As in the case of Pt or Ni doping, superconductivity can be easily suppressed, which may be ascribed to the positive internal pressure effect corresponding to the shrinking of the $c$-axis by the isovalent substitution of Pd, in spite of the enhanced SOC in the Pt case. It should also
be noted that superconductivity can survive up to very high doping
level in the cases of Ru or Ir doping despite the disorder, which demonstrates that
$T_c$ is quite robust in the hole-type doping, as
compared with the iron-based superconductors
LaFe$_{1-x}$Co$_x$AsO\cite{LaFe$_{1-x}$Co$_x$AsO} and
BaFe$_{2-x}$Ni$_x$As$_2$\cite{BaFe$_2$As$_2$}. Careful studies on
the single-crystalline Nb$_{2}$Pd$_x$S$_{5-\delta}$ revealed
that superconductivity occurs in a wide range of Pd ($0.6 < \emph
x < 1$) and S ($0 < \delta < 0.61$) contents\cite{fiber},
suggesting again that superconductivity in this system is very
robust.

Since a previous study proposed that changes in the SOC may affect the
large upper critical field ($H_{c2}$)\cite{Ni doping}, it is very
helpful to compare the ratio of $H_{c2}$ to $T_c$ for different
doping cases. We measured the temperature-dependent resistivity
for applied magnetic fields up to 15 T to get the upper critical
field data \emph H$_{c2}$ for the two series of samples
Nb$_{2}$Pd$_{1-x}$Ru$_{x}$S$_{5}$ and
Nb$_{2}$Pd$_{1-x}$Ir$_{x}$S$_{5}$,  as shown in Fig. \ref{fig5}.
The resultant \emph H$_{c2}$ data are summarized in Fig. \ref{fig5}(a)
and \ref{fig5}(b) together with the theoretical Ginzberg--Landau
(GL) fits (solid line)\cite{spin-orbit}. The \emph H$_{c2}$ values
were deduced from the fields at which $\rho$(\emph T) drops to
90$\%$ of the normal state value and the data can be well
described by the GL equation. In Fig. \ref{fig5}(c) and \ref{fig5}(d),
the variations in extrapolated \emph H$_{c2}$(0) values with the
doping content are displayed. For the
Nb$_{2}$Pd$_{1-x}$Ru$_{x}$S$_{5}$ samples, \emph \emph H$_{c2}$(0)
is initially enhanced and then goes down with the Ru substitution.
In the case of Ir doping, \emph H$_{c2}$(0) is revealed to
monotonically increase with increasing doping. In addition to the \emph
H$_{c2}$(0) data, \emph T$_c$ is shown for
comparison. Roughly speaking, the tendency of \emph H$_{c2}$(0) is
in good agreement with \emph T$_c$.  Compared with the calculated
\emph H$_{c2}$(0) data using the single-band
Wethamer--Helfand--Hohenberg(WHH) model $H_{c2}(0)=
-0.69 T_c$(d$H_{c2}$/d$T$)$_{T_c}$, the difference in the
$H_{c2}(0)$ values is below 5\%. In Fig. \ref{fig5}(e), a phase
diagram for \emph H$_{c2}$/$\emph T_c$ with respect to variation
of doping content is presented. Upon increasing the doping level,
$H_{c2}$/$\emph T_c$ exhibits a significant enhancement in this
system, exceeding the Pauli paramagnetic limit value (1.84$\emph
T_c$) by a factor of 3.8 to 4.6. This behavior is in support of the
fact that hole-type doping benefits superconductivity and thus
gives rise to larger upper critical fields. Moreover, a greater
enhancement of \emph H$_{c2}$/$\emph T_c$ in
Nb$_{2}$Pd$_{1-x}$Ir$_{x}$S$_{5}$ series than
Nb$_{2}$Pd$_{1-x}$Ru$_{x}$S$_{5}$ is observed, which may originate
from an additional SOC effect on the one-dimensional Pd chains by Ir
doping, in the framework of the scenario in which high superconducting
upper critical field may arise from strong SOC, as evidenced from
the opposite \emph H$_{c2}$/$\emph T_c$ tendency in the Pt and Ni
doping cases, reported by Zhou et al.\cite{Ni doping} These
findings again suggest that charge carrier as well as the large SOC in
this system have significant effects on superconductivity. It is
noted that \emph H$_{c2}$/$\emph T_c$ at \emph x = 0.4 for Ru
doping goes down slightly. The suppression could be ascribed to
impurity effects at a high doping level.

\section{Conclusion}
In conclusion, we compared the effects of Ru and Ir doping on
superconductivity in the Nb$_{2}$PdS$_{5}$ system. The enhancement
in both \emph T$_c$ and \emph H$_{c2}$ is observed upon partial Ru
(or Ir) substitution and the hole-type dominant charge transport
property are confirmed by Hall coefficient measurements.
However, the increases in $T_c$ and the ratio $H_{c2}$/$T_c$ is
more significant in the Ir doping case than in the Ru doping case.
Given that SOC in the system could be hardly changed by Ru
doping, but enhanced by Ir doping, the comparison of the two
doping cases suggests that there is a correlation between the SOC and
the enhanced $H_{c2}$/$T_c$ ratio. Our work reveals that the
exotic superconductivity in this system could be related to the
strong SOC on the Pd site.

The authors would like to thank Guanghan Cao for helpful
discussions. This work is supported by the Ministry of Science and
Technology of China (Grant Nos. 2014CB921203 and 2016YFA0300402),
NSF of China (Contract Nos. U1332209 and 11190023), the Ministry
of Education of China (Contract No. 2015KF07), and the Fundamental
Research Funds for the Central Universities of China.

\end{document}